# Considerations and Recommendations from the ISMRM Diffusion Study Group for preclinical diffusion MRI: Part 1 — In vivo small-animal imaging

Ileana O Jelescu[1,2,#], Francesco Grussu[3,4], Andrada Ianus[5], Brian Hansen[6], Rachel L C Barrett[7,8], Manisha Aggarwal[9], Stijn Michielse[10], Fatima Nasrallah[11], Warda Syeda[12], Nian Wang[13,14], Jelle Veraart[15], Alard Roebroeck[16], Andrew F Bagdasarian[17,18], Cornelius Eichner[19], Farshid Sepehrband[20], Jan Zimmermann[21], Lucas Soustelle[22], Christien Bowman[23,24], Benjamin C Tendler[25], Andreea Hertanu[1], Ben Jeurissen[26,27], Marleen Verhoye[23,24], Lucio Frydman[28], Yohan van de Looij[29], David Hike[17,18], Jeff F Dunn[30,31,32], Karla Miller[33], Bennett A Landman[34], Noam Shemesh[5], Adam Anderson[36,35], Emilie McKinnon[37], Shawna Farquharson[38], Flavio Dell' Acqua[39], Carlo Pierpaoli[40], Ivana Drobnjak[41], Alexander Leemans[42], Kevin D Harkins[43,35,44], Maxime Descoteaux[45,46], Duan Xu[47], Hao Huang[48,49], Mathieu D Santin[50,51], Samuel C. Grant[17,18], Andre Obenaus[52,53], Gene S Kim[54], Dan Wu[55], Denis Le Bihan[56,57], Stephen J Blackband[58,59,60], Luisa Ciobanu[61], Els Fieremans[62], Ruiliang Bai[63,64], Trygve B Leergaard[65], Jiangyang Zhang[66], Tim B Dyrby[67,68], G Allan Johnson[69,70], Julien Cohen-Adad[71,72,73], Matthew D Budde[74,75], Kurt G Schilling[43,35,#]

[#]Corresponding authors — ileana.jelescu@chuv.ch, kurt.g.schilling.1@vumc.org

[1]Department of Radiology, Lausanne University Hospital and University of Lausanne, Lausanne, Switzerland, [2]CIBM Center for Biomedical Imaging, Ecole Polytechnique Fédérale de Lausanne, Lausanne, Switzerland, [3]Radiomics Group, Vall d'Hebron Institute of Oncology, Vall d'Hebron Barcelona Hospital Campus, Barcelona, Spain, [4]Queen Square MS Centre, Queen Square Institute of Neurology, Faculty of Brain Sciences, University College London, London, UK, [5]Champalimaud Research, Champalimaud Foundation, Lisbon, Portugal, [6]Center of Functionally Integrative Neuroscience, Aarhus University, Aarhus, Denmark, [7]Department of Neuroimaging, Institute of Psychiatry, Psychology and Neuroscience, King's College London, London, UK, [8]NatBrainLab, Department of Forensics and Neurodevelopmental Sciences, Institute of Psychiatry, Psychology and Neuroscience, King's College London, London, UK, [9]Russell H. Morgan Department of Radiology and Radiological Science, Johns Hopkins University School of Medicine, Baltimore, MD, USA, [10]Department of Neurosurgery, School for Mental Health and Neuroscience (MHeNS), Maastricht University Medical Center, Maastricht, The Netherlands, [11]The Queensland Brain Institute, The University of Queensland, Queensland, Australia, [12]Melbourne Neuropsychiatry Centre, The University of Melbourne, Parkville, Victoria, Australia, [13]Department of Radiology and Imaging Sciences, Indiana University, IN, USA, [14]Stark Neurosciences Research Institute, Indiana University School of Medicine, IN, USA, [15]Center for Biomedical Imaging, NYU Grossman School of Medicine, New York, NY, USA, [16]Faculty of psychology and Neuroscience, Maastricht University, Maastricht, Netherlands, [17]Department of Chemical & Biomedical Engineering, FAMU-FSU College of Engineering, Florida State University, Tallahassee, FL, USA, [18]Center for Interdisciplinary Magnetic Resonance, National HIgh Magnetic Field Laboratory, Tallahassee, FL, USA, [19]Department of Neuropsychology, Max Planck Institute for Human Cognitive and Brain Sciences, Leipzig, Germany, [20]USC Stevens Neuroimaging and Informatics Institute, Keck School of Medicine of USC, University of Southern California, Los Angeles, CA, USA, [21]Department of Neuroscience, Center for Magnetic Resonance Research, University of Minnesota, MN, USA, [22]Aix Marseille Univ, CNRS, CRMBM, Marseille, France, [23]Bio-Imaging Lab, Faculty of Pharmaceutical, Biomedical and Veterinary Sciences, University of Antwerp, Antwerp, Belgium, [24]μNEURO Research Centre of Excellence, University of Antwerp,

Antwerp, Belgium, [25]Wellcome Centre for Integrative Neuroimaging, FMRIB, Nuffield Department of Clinical Neurosciences, University of Oxford, United Kingdom, [26]imec Vision Lab, Dept. of Physics, University of Antwerp, Belgium, [27]Lab for Equilibrium Investigations and Aerospace, Dept. of Physics, University of Antwerp, Belgium, [28]Department of Chemical and Biological Physics, Weizmann Institute of Science, Rehovot, Israel, [29]Division of Child Development & Growth, Department of Pediatrics, Gynaecology & Obstetrics, School of Medicine, Université de Genève, Genève, Switzerland, [30]Department of Radiology, Cumming School of Medicine, University of Calgary, Calgary, Alberta, Canada, [31]Hotchkiss Brain Institute, Cumming School of Medicine, University of Calgary, Calgary, Alberta, Canada, [32]Alberta Children's Hospital Research Institute, Cumming School of Medicine, University of Calgary, Calgary, Alberta, Canada, [33]FMRIB Centre, Wellcome Centre for Integrative Neuroimaging, Nuffield Department of Clinical Neurosciences, University of Oxford, Oxford, United Kingdom, [34]Department of Electrical and Computer Engineering, Vanderbilt University,, [35]Vanderbilt University Institute of Imaging Science, Vanderbilt University, Nashville, TN, [36]Department of Radiology and Radiological Sciences, Vanderbilt University Medical Center, Nashville, TN, USA, [37]Medical University of South Carolina, Charleston, SC, USA, [38]National Imaging Facility, The University of Queensland, Brisbane, Australia, [39]Department of Forensic and Neurodevelopmental Sciences, King's College London, London, UK, [40]Laboratory on Quantitative Medical imaging, NIBIB, National Institutes of Health, Bethesda, MD, USA, [41]Department of Computer Science, University College London, London, UK, [42]PROVIDI Lab, Image Sciences Institute, University Medical Center Utrecht, The Netherlands, [43]Radiology and Radiological Sciences, Vanderbilt University Medical Center, Nashville, TN, USA, [44]Biomedical Engineering, Vanderbilt University, Nashville, TN, [45]Sherbrooke Connectivity Imaing Lab (SCIL), Computer Science department, Université de Sherbrooke, [46]Imeka Solutions, [47]Department of Radiology and Biomedical Imaging, University of California San Francisco, CA, USA, [48]Department of Radiology, Perelman School of Medicine, University of Pennsylvania, Philadelphia, PA, USA, [49]Department of Radiology, Children's Hospital of Philadelphia, Philadelphia, PA, USA, [50]Centre for NeuroImaging Research (CENIR), Inserm U 1127, CNRS UMR 7225, Sorbonne Université, Paris, France, [51]Paris Brain Institute, Paris, France, [52]Div. of Biomedical Sciences, University of California Riverside, Riverside CA USA, [53]Preclinical and Translational Imaging Center, University of California Irvine, Irvine CA USA, [54]Department of Radiology, Weill Cornell Medical College, New York, NY, USA, [55]Key Laboratory for Biomedical Engineering of Ministry of Education, College of Biomedical Engineering & Instrument Science, Zhejiang University, Hangzhou, China, [56]CEA, DRF, JOLIOT, NeuroSpin, Gif-sur-Yvette, France, [57]Université Paris-Saclay, Gif-sur-Yvette, France, [58]Department of Neuroscience, University of Florida, Gainesville, FL, United States, [59]McKnight Brain Institute, University of Florida, Gainesville, FL, United States, [60]National High Magnetic Field Laboratory, Tallahassee, FL, United States, [61]NeuroSpin, UMR CEA/CNRS 9027, Paris-Saclay University, Gif-sur-Yvette, France, [62]Department of Radiology, New York University Grossman School of Medicine, New York, NY, USA, [63]Interdisciplinary Institute of Neuroscience and Technology, School of Medicine, Zhejiang University, Hangzhou, China, [64]Frontier Center of Brain Science and Brain-machine Integration, Zhejiang University, [65]Department of Molecular Biology, Institute of Basic Medical Sciences, University of Oslo, Norway, [66]Department of Radiology, New York University School of Medicine, NY, NY, USA, [67]Danish Research Centre for Magnetic Resonance, Centre for Functional and Diagnostic Imaging and Research, Copenhagen University Hospital Amager & Hvidovre, Hvidovre, Denmark, [68]Department of Applied Mathematics and Computer Science, Technical University of Denmark, Kongens Lyngby, Denmark, [69]Duke Center for In Vivo Microscopy, Department of Radiology, Duke University, Durham, North Carolina, [70]Department of Biomedical Engineering, Duke University, Durham, North Carolina, [71]NeuroPoly Lab, Institute of Biomedical Engineering, Polytechnique Montreal, Montreal, QC, Canada, [72]Functional Neuroimaging Unit, CRIUGM, University of Montreal, Montreal, QC, Canada, [73]Mila - Quebec AI Institute, Montreal, QC, Canada, [74]Department of Neurosurgery, Medical College of Wisconsin, Milwaukee, Wisconsin, [75]Clement J Zablocki VA Medical Center, Milwaukee, Wisconsin

## Abstract

Small-animal diffusion MRI (dMRI) has been used for methodological development and validation, characterizing the biological basis of diffusion phenomena, and comparative anatomy. The steps from animal setup and monitoring, to acquisition, analysis, and interpretation are complex, with many decisions that may ultimately affect what questions can be answered using the resultant data. This work aims to present selected recommendations and guidelines from the diffusion community, on best practices for preclinical dMRI of *in vivo* animals. We describe the general considerations and foundational knowledge that must be considered when designing experiments. We briefly describe differences in animal species and disease models and discuss why some may be more or less appropriate for different studies. We then give guidelines for *in vivo* acquisition protocols, including decisions on hardware, animal preparation, and imaging sequences, followed by advice for data processing including pre-processing, model-fitting, and tractography. Finally, we provide an online resource which lists publicly available preclinical dMRI datasets and software packages, to promote responsible and reproducible research. In each section, we attempt to provide guides and recommendations, but also highlight areas for which no guidelines exist (and why), and where future work should focus. While we mainly cover the central nervous system (on which most preclinical dMRI studies are focused), we also provide, where possible and applicable, recommendations for other organs of interest. An overarching goal herein is to enhance the rigor and reproducibility of small animal dMRI acquisitions and analyses, and thereby advance biomedical knowledge.

## 1. Introduction

Diffusion MRI (dMRI) is a non-invasive technique that exploits the hindered or restricted diffusion of water molecules in biological tissue to extract information about tissue microstructure in both normal and abnormal states. In the brain, dMRI-derived microstructure and connectivity maps have often been described as non-invasive 'virtual histology' or 'virtual dissection' and have found applications widely used in neuroanatomy, developmental, cognitive, and systems neuroscience, neurology, and neuroevolution, for exploring neural architecture in healthy and disease afflicted brains.

Both *in vivo* dMRI studies of small animals and of *ex vivo* specimens derived from animal or human tissues, have greatly contributed to scientific knowledge. In this work, **small animal imaging refers to imaging living animal models, whereas *ex vivo* refers to perfused living tissue or fixed tissue - the latter are covered in separate papers, 'Part 2' and 'Part 3'**. Many influential works in dMRI were first performed in small animals or *ex vivo* samples. For example, the discovery of a dramatically decreased diffusivity in cerebral ischemia was first observed in a cat model[1], diffusion tensor imaging formalism was originally validated on vegetables, pork loin, and rabbit models[2,3], and diffusion anisotropy was first observed again in cat models[4]. Microstructural and tractography models of today are routinely validated against animal models, *ex vivo* scans, and subsequent histological analysis[4–7].

The added value of preclinical dMRI is multifold (**Figure 1**). First, small animal dMRI allows correlations with histological and other (invasive) imaging measures to discover the biophysical basis of the dMRI signal, parameters, and biomarkers with the ultimate goal to act as a non-invasive in vivo microscope. Second, small animal imaging allows acquisition of "extreme" datasets, at the edge of what is achievable in dMRI in terms of spatial resolution and/or diffusion-weighting coverage, and clearly beyond what is currently achievable with clinical imaging, leveraging the access to much stronger gradients, and to longer scan times. Together these allow the acquisition of more comprehensive datasets. Third, the use of animal models allows us to study the sensitivity of dMRI to tissue changes in diseases, disorders and treatments in a controlled way that is not always possible in humans, allowing the knowledge gain to be applied to human studies. Small animal MRI can be multimodal, easily longitudinal, and supported by behavior analysis and molecular techniques (optogenetics, omics, immunohistochemistry, etc.) from the same animals. Fourth, the use of animal models enables comparative anatomy, allowing the investigation of differences between human and other mammalian brains.

The science of dMRI covers many disciplines and is continually evolving. The steps from animal setup and monitoring, to acquisition, analysis, and interpretation are complex, with many decisions that may ultimately affect what questions can be answered using the data. **The present work does not serve as a "consensus" on any specific topic, but rather as a snapshot of "best practices" or "guidelines"** from the preclinical dMRI community as represented by the authors. Recruitment for participation in this effort included two meetings of the Diffusion Study Group of the International Society for Magnetic Resonance in Medicine, responses to a survey (*Supplementary Material*) distributed within the Diffusion Study Group forum, and recommendations for recruitment from other authors. We envision this work to be useful to imaging centers using small animal scanners for research, sites that may not have personnel with expert knowledge in diffusion, pharmaceutical or industry employees who may want to run their own tests and studies, or new trainees in the field of dMRI.

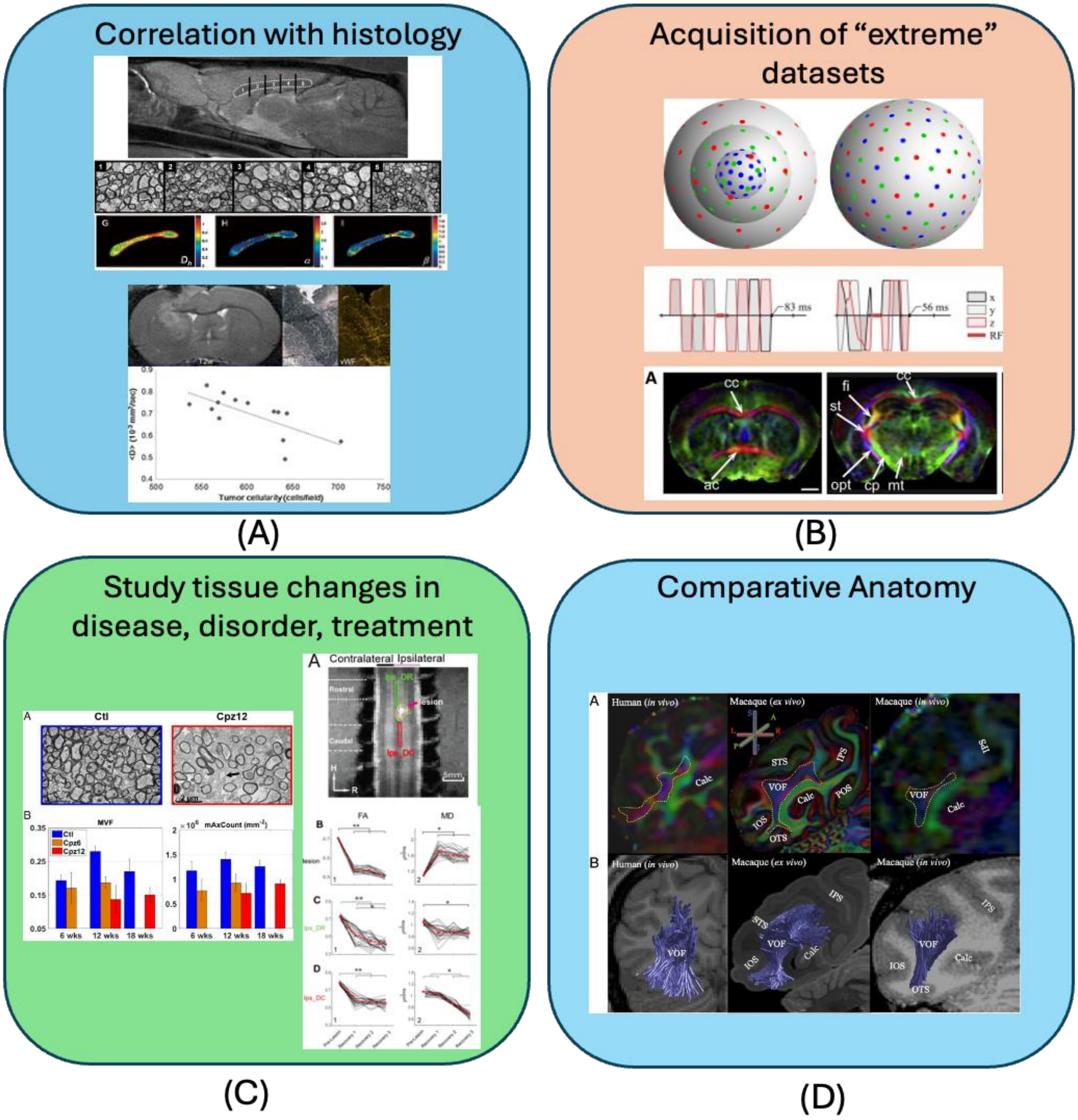

**Figure 1**. Four areas in which preclinical brain imaging adds value to the field of dMRI. It enables: (i) correlation with histology on the same subject/sample, (ii) the acquisition of richer datasets than on clinical systems thanks to more advanced hardware and longer scan times available, (iii) the study of tissue changes with disease and treatment in a more controlled setting, and (iv) comparative anatomy between species. Figures reused and adapted from (A) [17,18] (B) [19–22] (C) [23,24] (D) [25].

The manuscript is organized as follows. We first describe **general considerations and foundational knowledge that must be considered when designing experiments**. We briefly describe **differences in species and models** and discuss why some may be more or less appropriate for different studies. We then give guidelines for ***in vivo* acquisition protocols**, including decisions on hardware, animal preparation and imaging sequences, followed by recommendations for **data processing** including pre-processing, model-fitting, and tractography. Finally, we give **perspectives** on the field, describing **sharing of code and data**, and **goals** that we wish to achieve. We also highlight areas for which no guidelines exist (and why), and where future work should lie. An overarching goal herein is to **enhance the rigor and reproducibility** of small animal dMRI acquisitions and analyses, and thereby advance biomedical knowledge.

## 2. Translational aspects

### 2.1 Translation and validation considerations

Several aspects must be considered when designing and performing experiments to appropriately interpret scientific results, including. the tissue model itself, disease and disorders, and hardware and experimental setup. While the diffusion process is fundamentally the same, and governed by the same laws of physics, these must be carefully thought out to translate findings to the *in vivo* human.

Basic constituents of the brain and other organs are largely preserved across mammalian species, providing a basis for translational *in vivo* MRI studies[26,27]. In the **central nervous system**, the fundamental structure of axons with a myelin sheath makes the dMRI signal interpretation in white matter fundamentally translatable, although there are variations in axon diameter, myelin thickness and ratio of myelinated to unmyelinated axons[28,29]. Brain cortical layers are also largely preserved across species. However, the ratio of white-to-gray matter is very different between rodents and primates, with predominant gray matter, unfolded cortex and thin white matter tracts in rodents, and relatively more white matter and folded cortex in primates (see Mota et al.[30] for a comprehensive characterization across mammalian species). As a result, partial volume effects are more challenging to mitigate in rodent white matter and in human/primate gray matter, respectively (**Figure 2**). Moreover, the complexity of white matter organization in rodents is different from primates, resulting in potential issues when translating modeling and tractography approaches from the rodent brain. However, large bundles such as corpus callosum, the external capsule and the fimbria/fornix retain structure similarity, even in rodents.

For **other organs**, intrinsic differences in microstructure do exist, such as much larger hepatocytes in mouse liver compared to humans[31,32], composition of endocrine islets in the pancreas[33], the existence of a marginal zone in the mouse spleen[34], etc; thus potential differences need to be considered when interpreting results.

**Brain injury models**, such as traumatic brain injury (TBI), epilepsy, stroke, subarachnoid and intracerebral hemorrhage, spinal cord injury, edema or de/re-myelination have high translational value since the cellular responses to the external insults are similar between species [23,35–46]. **Tumor models** may also display some translational value [18,45–49]. Even if animal models are not directly translatable, they may offer partial systemic deficits that mimic relevant aspects of the disease (i.e. altered microstructure or connectivity) as in **models of neurodegenerative and psychiatric diseases**[23,50,51].

The difference in gradient strength and diffusion times between small animal and clinical acquisitions results in a different sensitivity to spatial scales, such as cell sizes, packing density, etc. Hence, in addition to careful selection of appropriate animal models relevant to the corresponding clinical situation, the diffusion acquisition parameters should be matched carefully in preclinical dMRI studies that aim to validate clinical results.

## 2.2 Species differences

Species have different pros and cons for imaging studies - which we briefly cover in this section. From an ethical standpoint, it is always recommended to go with the least evolved species that answers the research question at stake; in particular, NHPs should only be considered when the sought effect or structure is not present in rodents.

### 2.2.1 Murine models (mouse and rat)

Rats and mice have been the long-standing preferred species for biomedical research, including dMRI. Advantages include wide availability, group homogeneity, well-characterized transgenic models simulating human pathology, and very rapid lifespan. The ability, particularly in mice, to rapidly insert human genes allows great flexibility pursuing genomic functional changes that may mimic the human condition. Rats on the other hand are less challenging to image, due to larger structure sizes, which improves field homogeneity and thus image quality, and requires lower spatial resolution to resolve fine structures. In addition the small physical size offers technical advantages, fitting in the typically smaller bores (and smaller coils) of magnets with larger field strengths.

It is however important to note that multiple anatomical differences highlighted in Section 2.2, limit the direct translatability of dMRI findings from murine models to humans.

### 2.2.2 Primate models

Non-human primates (NHPs) including marmosets, squirrel monkeys, and macaques. They are commonly used in neuroscience research because their brain has a large number of white matter and gray matter regions with homologous human counterparts. NHPs are thus well suited for studies of cortical development, gyrification, and interrogation of complex white matter. NHPs allow access to “ground truth” connectivity, which has been well-documented through the use of tracer and ablation studies[53].

Controversies regarding the existence or nonexistence of a pathway, or the location of pathway terminations have been resolved through primate models[54–57], and a number of tractography validation studies have used tracer studies in primates[58–61], although validation is most commonly performed on *ex vivo* samples prior to histological analysis[62–71].

Primate species must also be considered for each experiment: smaller monkeys (galago, squirrel monkey) may be easier to work with, less cumbersome to scan, less expensive to house, and the reduced gyrification makes cortical identification easier (for example for injections or electrical stimulation). The disadvantages of NHPs are often associated with access costs. Complex housing is required, training for transportation to the scanner and preparation for scanning. Lastly, small bore preclinical systems and high performing gradient inserts are often not large enough for bigger brains.

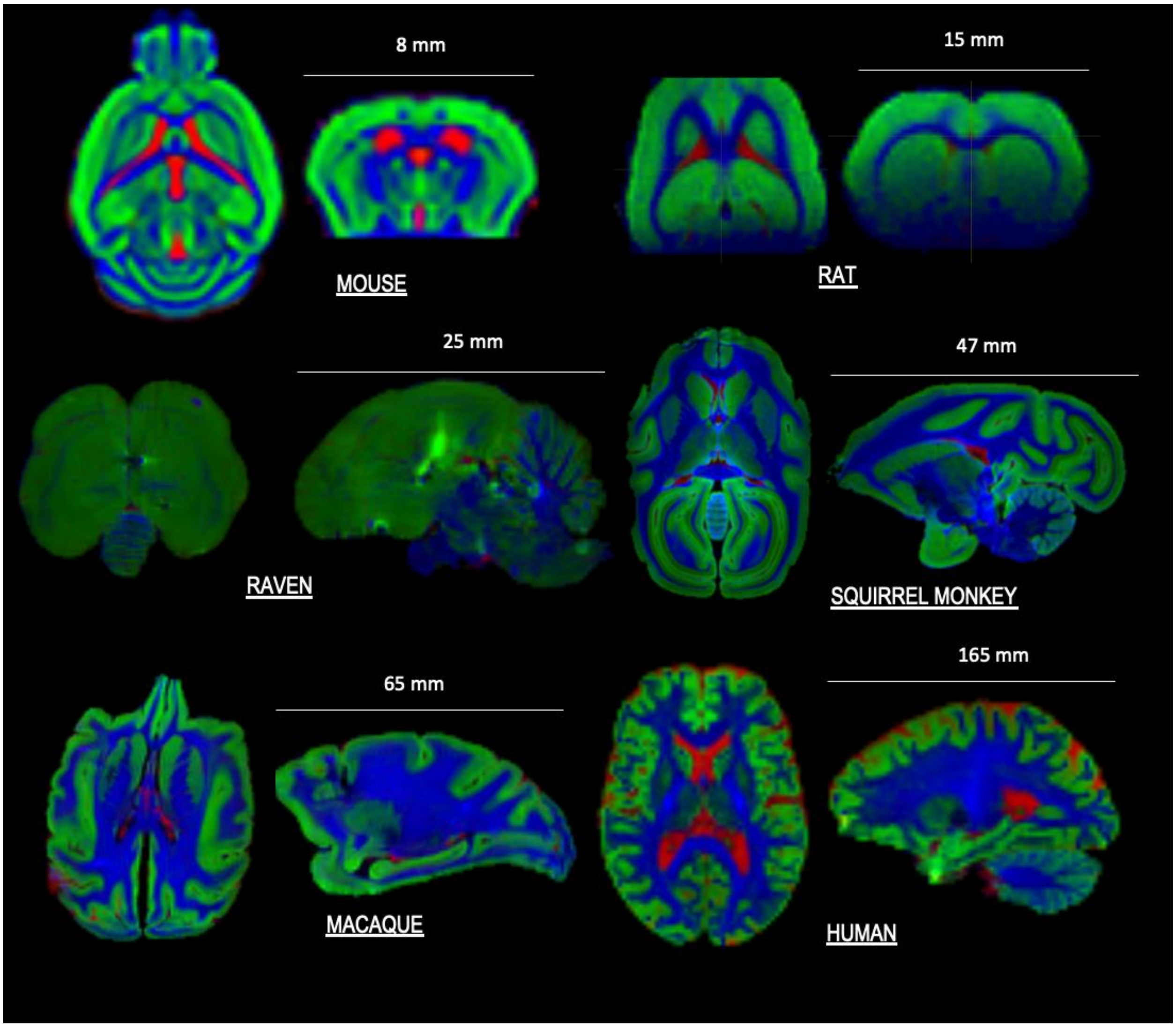

**Figure 2**. dMRI brain images of small animal models demonstrating different brain sizes, geometric complexity, gyrification, and tissue constituents, ordered by increasing complexity. Different tissue types are estimated using Multi-Shell Multi-Tissue Spherical Deconvolution[52] and color coded - CSF (red), GM (Green), and WM (blue). *In vivo* data: mouse, rat, human. *Ex vivo* data: raven, squirrel monkey, macaque. Data kindly provided by Adam Anderson, Ileana Jelescu, Kurt Schilling, Ben Jeurissen and Marleen Verhoye.

2.2.3 Other models

While murine and NHP models are most widely used in dMRI research, other models include the pig brain, which is comparable to the human brain in myelination and development[72,73], and has been used to study development[73], brain lesions[74] and tractography validation[71,72,75]. Other gyrencephalic brains (e.g. ferrets) have been used to study psychiatric diseases, cognition and brain function, or to validate tractography[42,76,77]. Songbirds have been used to study fundamental properties of naturally occurring neuroplasticity[78]. Of particular interest, diffusion anisotropy and stroke were first experimentally observed and demonstrated in cat models[1,4,79].

# 3. Acquisition

Important decisions when performing dMRI of small animals include selecting appropriate hardware, animal preparation and monitoring, and acquiring the data (**Figure 3**). In this Section, we present a **recommended 'standard' imaging setup and acquisition protocol that can be achieved in 20-30 minutes, and would be appropriate for a wide range of diffusion applications and analyses.** Longer scan times allow for richer dMRI data, and ultimately, the chosen protocol should be suitable for the planned analysis.

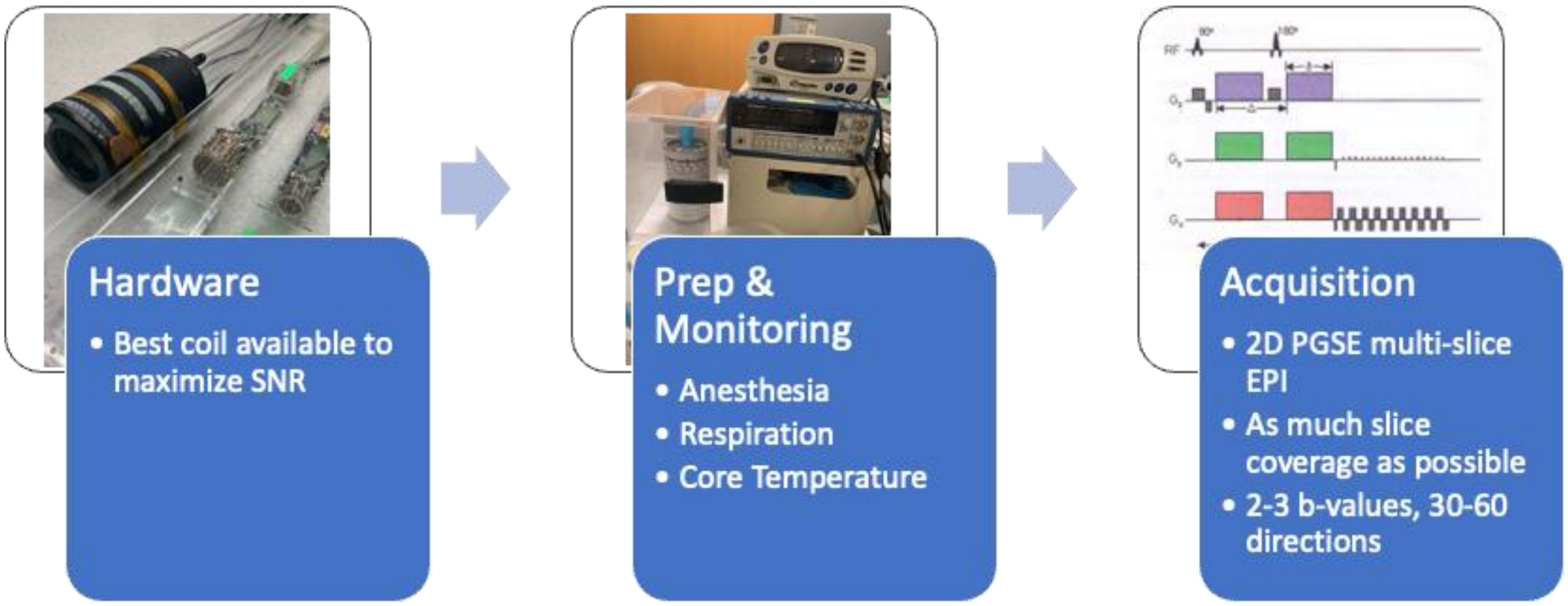

**Figure 3**. Small animal *in vivo* protocols require decisions regarding hardware, animal preparation and monitoring, and acquisition (which includes encoding, readout, spatial resolution, and *q-t* coverage).

It should be emphasized that these suggestions reflect a typical protocol as a starting point for many studies. Detailed information on each aspect is described to justify our recommendations and we highlight other strategies to optimize diffusion acquisition for a desired experiment.

## 3.1 Hardware (species/organ specific)

### 3.1.1 RF Coils

The key to maximizing SNR is using an appropriate RF coil with a high filling factor, defined as the magnetic field energy stored inside the sample volume versus the total magnetic energy[80,81]. To this end, combinations of a volume coil for excitation and a surface coil for reception are often used, as they ensure homogeneous excitation and maximum reception sensitivity. For small bore magnets, where the available space is limited, transceiver surface coils can be utilized. Surface coils perform well when the region of interest is located to the surface. For deeper regions, small volume transmit/receive coils can provide better sensitivity in addition to the improved $B_1^+$ homogeneity and are recommended for whole brain studies. Phased array coils cover a larger field of view and can be used in conjunction with parallel imaging strategies for acquisition speed-up[82]. The use of phased arrays is recommended for larger animals where one can afford to trade SNR for acquisition time reduction[83].

Cryogenic probes (cryoprobes) can increase SNR by factors of 2.5 to 5 compared to standard room-temperature RF coils, by minimizing thermal coil noise[84,85]. Most commercially available cryprobes are for mouse brain, and some for rat brain. Indeed, cryoprobes are most beneficial on small samples where the electronic noise dominates. Although cryoprobes operating at ultra-high field exist[86], the biggest gains are obtained at lower field strengths. In fact, the SNR gains provided by cryoprobes can bring the signal sensitivity performance of systems operating at 7 - 9.4 T to the levels achieved with ultra-high-field systems (≥11.7 T) without the penalty in relaxation times (shorter $T_2$ and longer $T_1$ at higher fields) and susceptibility artifacts.

Diffusion-weighted images are typically low SNR. The choice of RF coil is thus of critical importance for the quality of in vivo diffusion acquisitions. Recent work by multiple teams[87–89] has demonstrated the utility of providing open source hardware designs for radiofrequency coils and holders.

### 3.1.2 Gradients

Clinical systems are typically equipped with 40 - 80 mT/m magnetic field gradients while those on small animal MRI systems are often 300 mT/m or higher, with 1 T/m becoming increasingly widespread. Some small animal systems are available with a high power gradient option, and dedicated inserts with up to 3 T/m (along all three axes) are commercially available.

Strong field gradients (G) allow an independent or largely decorrelated exploration of the two dimensions of q-t space — where q is the spatial phase warp that introduces diffusion sensitization ($q = \gamma\delta G$) and t is the diffusion time in which the molecules can diffuse and explore the local environment. The combination of the two yields the “b-value” that quantifies overall diffusion weighting ($b = q^2 t$, in the narrow pulse approximation $\delta \to 0$). Strong field gradients with rapid switching times further benefit diffusion experiments by enabling fast readouts and short TE’s to compensate for shorter $T_2$ at higher field. Our recommendation is to select the MRI system equipped with the strongest and fastest gradient system that is appropriate for the size of the *in vivo* animal imaging setup.

Stronger gradients present several challenges, including calibration, gradient nonlinearities, and eddy currents. Gradients must be well-calibrated to ensure accurate gradient fields, and hence, accurate diffusion weightings. Similarly, gradient fields are typically linear at the center of the coil (isocenter), but may deviate at locations further away. The gradient non-linearity can be mapped and corrected for during diffusion quantification[90], to reduce its bias on diffusivity estimation particularly for large samples relative to the gradient dimensions. Finally, fast switching gradients induce currents in MRI hardware components, causing eddy current artifacts that must be compensated for, or corrected in processing.

**Where future work should lie**

Parallel imaging is still very limited on pre-clinical scanners due to the object's small size and thus the low number of receiver coil elements (usually 1 - 4 for brain imaging). This limitation is due to the reduced space around the animal which caps the number of pre-amplifiers that can fit in the coil. Acquisition with rodent brain array coils typically makes use of parallel imaging methods (e.g. GRAPPA[91]) but most often achieves less reduction in scan time than in clinical imaging where many more receive channels are available. Progress in this field would be valuable for reducing scan times and artifacts such as ghosting[92–94]. Fortunately, rapid advances in RF circuitry will likely lead to an increase in the number of channels for small animal scanners.

Obstacles to standardization in preclinical studies is the large variability in preclinical imaging instrumentation as compared to clinical MRI systems. Indeed, there is a broader range of possible field strengths ( 4.7T to 17.2T) and gradient capabilities ( 400 mT/m to 3 T/m inserts) that will ultimately affect the achievable protocol. While here we intend to provide recommendations and guidelines to avoid substantial pitfalls rather than aim at standardization, our community should work towards proposing standardized protocols (and processing pipelines) for preclinical settings that would be achievable on a wide range of systems. This could be achieved following the example of the QIBA (Quantitative Imaging Biomarkers Alliance) guidelines for dMRI in the clinic[95]. Preclinical dMRI standardization may help reproducibility and harmonization for diffusion metrics that are simple (e.g. DTI) and common for characterizing animal models of disease. Such preclinical acquisition guidelines have recently been proposed for rat functional MRI[96], for example.

## 3.2 Animal preparation & physiological monitoring

General considerations about experimental design from the biological perspective, planning the experimental protocol, details on equipment needed for anesthesia and monitoring are beyond the scope of this review and the reader can refer to existing literature[97–100]. We underline however that maintaining stable physiological homeostasis during in vivo imaging is important. Reducing stress and physiological differences between animals helps to reduce variability which is particularly important for identifying group differences and for longitudinal studies.

A minimum monitoring setup should include respiration rate and core temperature. Both respiration and temperature sensors are connected to the physiological monitoring interface and software. Respiration rate is typically monitored using a pillow-sensor placed under the animal’s abdomen.  The temperature is monitored via a rectal thermometer and maintained using a warm water circulation system around the animal's body, or warm air blown into the MRI tunnel. A key factor for animal homeostasis is that maintenance is sufficient to prevent edema or brain swelling, and prevent whole body dehydration (a particular problem with small animals). A constant core temperature is also warranted to improve consistency of diffusivities and of $T_1$-weighting throughout the protocol, both of which are temperature-dependent[101]. A rule of thumb of ΔD = 0.06 $\mu m^2$/ms/°C for free water (interpolating between D=2 μm2/ms at 20°C and D=3 $\mu m^2$/ms at 37°C)[102], translates approximately into ΔD=0.02 $\mu m^2$/ms/°C in living tissue (assuming D~1 $\mu m^2$/ms at 37°C), consistent with [101].

**Anesthesia:** Isoflurane is often the anesthetic of choice given its ease of use and stability over long periods, but others can also be used [103]. It should be noted that a stable temperature and respiration rate of the animal are targeted, and thus the anesthesia level should be adapted to maintain that. For prolonged scans, the isoflurane concentration can be gradually reduced if the animal respiration rate decreases.

Data acquired under different anesthetic conditions (anesthesia types, dosage, effective breathing rate, etc.) should be compared with caution, especially as the literature is conflicted[103]. dMRI studies that examine cellular or tissue level effects typically have limited dependence on the anesthetic agent. However, in the mouse brain, mean diffusivity (MD) and mean kurtosis (MK) were both found to be lower under isoflurane than in the awake state[104], potentially due to inhalation isoflurane decreasing the brain extracellular space volume (cell swelling)[105]. In rats, brain ADC has been reported to increase with increasing anesthetic agent dosage for both isoflurane and medetomidine[106]. Furthermore, dMRI used to probe physiological effects such as flow or exchange may require more consideration of the effects of each agent on physiology (vasodilation/constriction, hypoxia and/or temperature), that have been well-established in the context of functional MRI, for example[104,105].

Recently, advantages in the physiological properties of dexmedetomidine over its racemate medetomidine have been reported and large animal laboratories have started converting their anesthetic protocols[110,111]. As effects also vary with the duration of anesthesia, it is recommended to keep timings consistent within an experimental cohort, and thus the delay between anesthesia onset and start of the dMRI acquisition.

For post-scan animal recovery, inhaled anesthetics (e.g. isoflurane) have faster elimination through the lungs while injectable ones need to be metabolized and excreted (e.g. medetomidine). An antagonist to the latter is recommended to speed up the recovery phase for the animal (e.g. atipamezole).

Positioning: Prone is the most common positioning for the animal. However, in the case of spine imaging, the supine position may be preferred to minimize organ-induced motion of the spine as well as reduce the distance between the spine and the coil. For liver imaging, the animal may be placed on its flank.

**Physiological Gating:** Brain dMRI does not typically require respiratory or cardiac gating provided the head is sufficiently stabilized within an appropriate holder which for rodents includes both teeth and ear bars. For organs susceptible to motion, gating strategies to limit the effects of motion are typically imperative. Respiratory and cardiac gating are available on most MRI systems[112–115]. Prospective gating is the most frequently used method that acquires data intermittently in response to an external trigger, which serves to minimize artifacts in body or spinal cord dMRI or to obtain images from the same phase of motion (e.g. in cardiac dMRI). Retrospective gating techniques that reorganize data after a continuous acquisition are not routinely used for dMRI. Gating typically prolongs acquisition times but improves the quality of the resulting images. Radial sampling strategies instead of EPI are also an alternative to gating, as they reduce respiration artifacts (ghosting) in abdominal imaging for example - but they also come at the expense of reduced SNR[116].

**Where future work should lie**

Systematic reporting of animal monitoring and anesthesia procedures as well as the resulting physiological measures (e.g. respiration rate and core temperature) in dMRI publications can contribute to improved reproducibility and multi-site comparison of results.

### 3.3 Diffusion encoding

A number of possible diffusion encoding, or sensitization, schemes are shown in **Figure 4** (left)[118] . Most diffusion-weighted sequences trace their origin to the **pulsed gradient spin-echo (PGSE)**[119] encoding scheme pioneered by Ed Stejskal and John Tanner in 1965. PGSE offers a mathematically elegant way to quantify diffusivity, gives access to a biologically relevant range of diffusion times (e.g. 10-50 ms on a preclinical MRI system at high field) and a broad range of *b*-values (e.g., 0—10’000 s/mm$^2$). For this reason, PGSE has become the most widely used diffusion encoding strategy in both human and animal imaging, and is the “default” encoding scheme on all current scanners.

Alternative diffusion encodings are possible, particularly on preclinical systems. Stronger gradients can take a variety of shapes, enabling sensitivity to different microstructural features, such as microscopic FA using multi-dimensional diffusion encoding (MDE), or short length scales using oscillating gradients. While there is no “consensus” on the best encoding strategy, we describe their pro’s and con’s, and when possible, suggest guidelines when using these sequences.

To probe long diffusion times (e.g. > 80 ms) a **stimulated echo acquisition mode (STEAM)** sequence may be useful, where signal recovery is limited by $T_1$ recovery rather than by $T_2$ decay[120]. The downsides of STEAM is a two-fold loss in SNR compared to PGSE (all other factors being equal) and larger contributions from imaging gradients to the diffusion-weighting via cross-terms. The latter aspect makes the calculation of the effective *b*-value mandatory as it can differ substantially from its nominal value, and can raise potential issues with large variations in effective *b*-values across directions over the same “shell”[121].

**Oscillating gradient spin echo (OGSE)**[122] can be used to probe much shorter time and length scales (0.1-10ms). This comes at the cost of modest attainable *b*-values, and a higher risk of nerve stimulation due to rapidly switching strong gradients. The *b*-value can be increased by lengthening the duration of the oscillating waveforms, enabling short diffusion times with moderate diffusion weightings.

In the above-mentioned schemes, the signal is sensitized to diffusion along a *single* gradient direction, often referred to as a **single diffusion encoding (SDE)** experiment. **MDE** techniques encode diffusion by using multiple sets of diffusion encoding gradients along (possibly) different directions. MDE examples include double diffusion encoding (DDE)[123,124] or triple diffusion encoding[125,126]. This increases the dimensionality of possible controllable parameters, and enables probing microscopic anisotropy[127,128], compartmental kurtosis[129], or compartmental exchange[130–132]. Further, free gradient waveforms may be designed for investigations into diffusion microscopic anisotropy, structure size variance, and orientational coherence, and is often referred to as *q*-trajectory imaging (QTI)[133]. Disadvantages of these sequences include potentially longer acquisition times, long echo time, ill-defined diffusion times, or complicated modeling and analysis. For a review of multidimensional diffusion encoding see[126,134].

Because of the strong preclinical gradients, eddy currents can create artifacts in diffusion images. It is possible to modify the diffusion encoding to a **twice-refocused spin echo**. This eddy current nulled (ECN) encoding can minimize these artifacts, although at the cost of decreased encoding efficiency and a longer TE.

There are many ways to encode diffusion into MRI images. With tradeoffs in diffusion times, diffusion weightings, sequence time, microstructure sensitivity, and artifact sensitivity, the optimal encoding strategy ultimately depends on the experimental question and desired application.

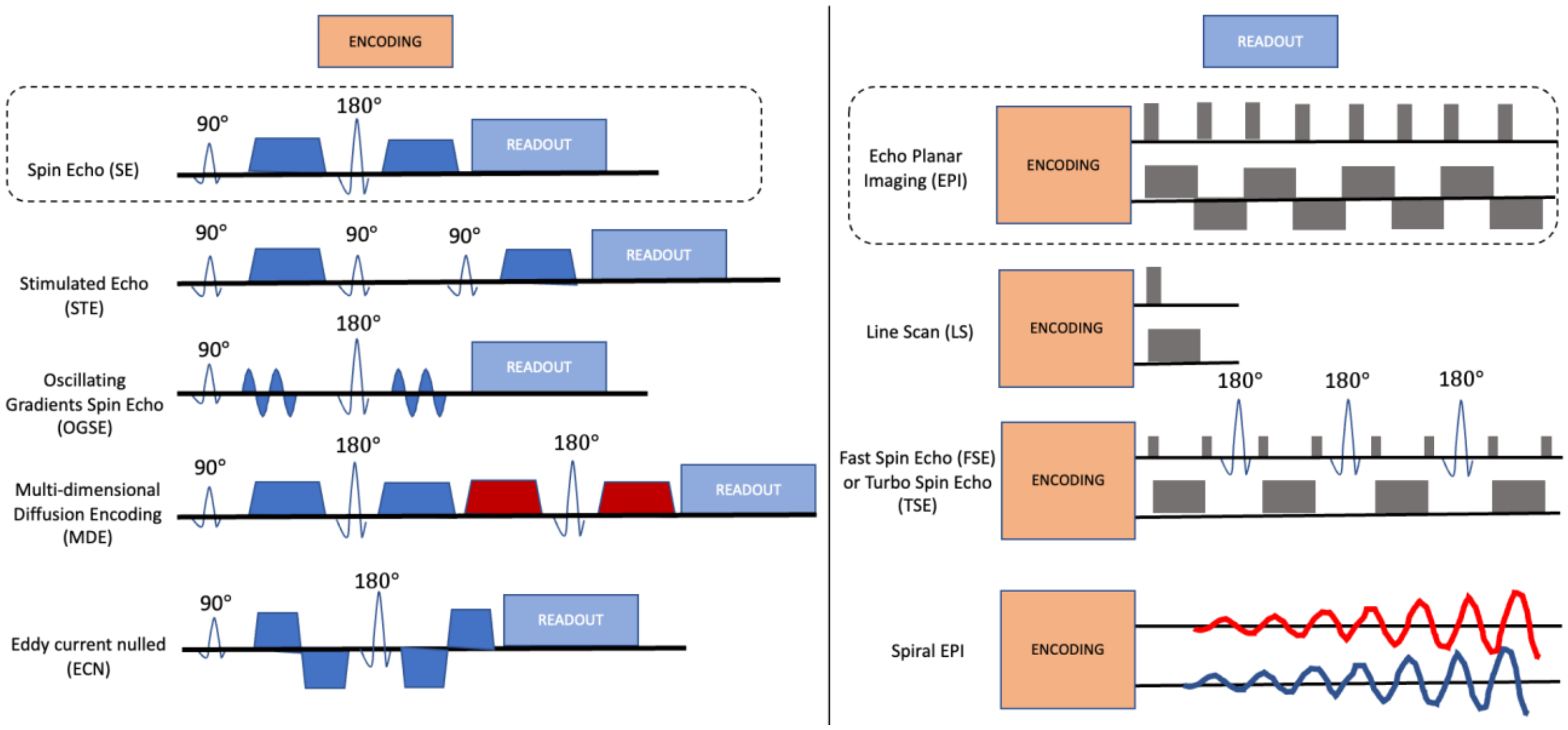

**Figure 4**. Diffusion encoding (left) and readout (right). Pulse sequence diagrams are shown for a variety of representative encoding (Section 3.3) and Readout (Section 3.4) schemes. Left: RF pulses are represented as hollow waveforms, diffusion gradients as dark color filled shapes (colors represent encoding axes), the readout module as pale blue. Slice selection gradients are not shown, for simplicity. Right: RF pulses are represented as hollow waveforms, gradients are in dark gray or thick lines. Two axes are shown for phase-encode and read-out gradient directions.

### 3.4 Signal readout

After diffusion encoding, the **signal readout** module is played out (**Figure 4**, right). **Echo planar imaging (EPI)** readout is a rapid acquisition that minimizes effects of bulk motion on diffusion images, and acquires images in a single excitation to minimize scan time needed to acquire many diffusion volumes. The EPI readout is also compatible with all diffusion encodings. A **single-shot multi-slice pulsed-gradient EPI sequence** has thus become the most popular in humans, and is also our recommended starting point for small animal *in vivo* imaging. Depending on hardware performance, multi-shot instead of single-shot EPI or a smaller imaging matrix can be implemented, until the image quality becomes acceptable in terms of spatial distortions.

While EPI reduces scan time and motion artifacts, it also faces its own challenges, most prominently spatial distortion due to $B_0$ field inhomogeneity that can be especially challenging at high field strengths, typical of preclinical imaging. To alleviate this, it is possible to acquire data with a segmented readout (multi-shot, as above), at the cost of increased scan time and possible ghosting artifacts, or use partial Fourier acceleration in the phase direction, at the expense of SNR and image sharpness. Finally, acquiring additional reversed phase-encoded $b$=0 images can help in compensating for geometrical distortions during image preprocessing (see Section 4.1.1).

Other readouts are also possible for small animal *in vivo* imaging. For example, **rapid acquisition with relaxation enhancement (RARE)**, sometimes called **fast spin echo (FSE) or turbo spin echo (TSE),** is less prone to susceptibility induced artifacts[135,136], or **gradient and spin echo (GRASE)** which also allows acquisition acceleration[136]. Additionally, *k*-space can be sampled in a **spiral readout** rather than cartesian readout. Spiral acquisitions have several advantages including the possibility to reduce TE. This readout remains nonetheless sensitive to $B_0$ field inhomogeneities and eddy currents (yielding geometric distortions) in the higher spatial frequencies. While this complicates artifact correction after image reconstruction, eddy currents can be minimized via trajectory measurements[137].

Other approaches combine diffusion weighting with relaxometry measurements in a multi-dimensional acquisition, for instance using multiple gradient echoes or spin echoes. Such sequences as well as the respective data reconstruction/analysis pipelines are provided in the REMMI toolbox (https://remmi-toolbox.github.io/) for different vendors.

### 3.5 q-t coverage

Unique insights into diffusion signal behavior, and thus into the underlying microstructure, have been brought by exploring a range of q-t regimes on animal systems, including very short diffusion times[8–11], very strong diffusion weighting[12,13] or more complex diffusion encoding schemes[14–16].

The “optimal” *q-t* coverage is highly dependent on the microstructure feature one wishes to maximize sensitivity to, therefore there is no direct recommendation on this topic[138]. It is important to determine upfront the theoretical requirements of the data analysis framework which will be used downstream, such as short gradient pulses (narrow pulse approximation), short / long diffusion times, *b*-value regime, etc. In general, on preclinical systems there is a lot of flexibility in choosing the sequence parameters which is very handy for optimisation strategies. We recommend acquiring non-diffusion weighted images (*b*=0) at approximately a ratio of one image for each 10-20 diffusion-weighted images[139,140].

**DTI** is by far the most widespread analysis of dMRI data. To maximize precision, the *b*-value should be chosen such that the signal decay is substantial, e.g. such that $bD \simeq 1$. *In vivo*, the most standard *b*-value is *b* = 1000 s/mm$^2$, as in humans. The b-value can be adapted based on the animal's body temperature, e.g. songbirds have a higher body temperature (40-41°C), leading to the use of lower *b*-values for DTI.

**DKI** is an extension of DTI that also estimates non-Gaussian characteristics of water diffusion in  biological tissues (diffusion kurtosis) from at least two non-zero shells with a recommended 20-30 directions per shell[141]. The highest *b*-value should be chosen as $b \simeq 2000 - 2500\, s/mm^2$ *in vivo*. Often the rule of thumb $b_{max} < 3/DK$ can be used, as from this point on the DKI signal expression starts to increase with increasing *b*. Given recent advances in acquisition efficiency and speed, we advocate a minimal acquisition of two-shell data leading to DKI analysis as the new “default” over simple DTI derived from single-shell data. A two-shell acquisition also opens up data analysis possibilities beyond DKI, e.g. for multi-compartment biophysical models[7]. Note, however, that optimal two-shell design may depend on the target tissue[142]. For this reason, it may be beneficial to use highly sampled repository datasets for study planning[143]. For reduction in DKI scan time, strategies for “fast kurtosis” estimation decrease the minimal number of measurements needed[144,145] assuming axial symmetry of diffusion properties in the voxel[146]. DKI and DTI rely on an empirical representation of the diffusion signal and make no assumptions about the underlying microstructure or tissue properties[6] and are thus widely applicable.

Guidelines for *q-t* **tractography**: Current tractography methods are less sensitive to scan parameters (*b*-values, number of gradient directions) than choices in the tractography pipeline itself (i.e., fiber orientation reconstruction, seeding strategies, streamline propagations, etc.)[64,69,147–151]. Our recommended protocol includes 50-60 directions at a moderate-to-high *b*-value[152]. This protocol is also compatible with tools such as ‘tractogram filtering’ and ‘microstructure-informed’ tractography[153–158] which are expected to increase anatomical accuracy of tractography. Note a Cartesian sampling of q-space, known as Diffusion Spectrum Imaging (DSI) can also be used[159,160].

Guidelines for acquisition in the perspective of **compartment modeling** analysis[7,161,162] are generally the same as in humans, with one nuance: as mentioned, preclinical scanner hardware may allow the exploration of regions of *q-t* space that are not achievable on clinical systems. Thus, the data requirements of some biophysical models may be better met for small animal imaging than human imaging.

Shorter **diffusion times** are typically favored or enforced when special effort is put into minimizing the echo time. These differences may pose additional challenges when extrapolating results obtained in small animals to humans. We therefore underline the *importance of reporting the diffusion time as part of the acquisition parameters*, even when the diffusion time dependency is not the focus of the study. As mentioned earlier, diffusion metrics measured at typically short diffusion times (10 - 20 ms) in preclinical experiments can be different to those measured in clinical studies with substantially longer diffusion times (typically longer than 50 ms) because they will be sensitive to different aspects and spatial scales of the micro-environment.

Preclinical scanners offer a unique opportunity to investigate diffusion **time-dependence** of diffusivity D(t) or kurtosis K(t) over diffusion time ranges that are difficult to achieve on clinical systems (in particular, short diffusion times either using OGSE or PGSE), concomitantly with the exploration of high *b*-values[8–10,163–167]. For studies with short to intermediate diffusion times (up to ~40 – 50 ms) we recommend to keep the echo time constant to a value which allows for the longest gradient separation (Δ in a PGSE scheme). Indeed, except for a combined diffusion-relaxometry model, accounting for variable $T_2$-weighting considerably complicates the data analysis. We underline this constant TE recommendation as it is commonly (and unfortunately) not the primary choice due to SNR considerations. For longer diffusion times (>80 ms), a STEAM sequence can be used[121,168].

Beyond the brain, organs such as the liver are highly vascularised and feature strong intra-voxel incoherent motion (**IVIM**) effects. The anisotropy of the dMRI signal is however much smaller than in the brain, so that directional schemes based on 3 mutually orthogonal diffusion directions can characterize tissue diffusion properties well in most cases[169]. Furthermore, cell sizes are much larger than in the brain (e.g., 20-40 μm for hepatocytes[32,170]), implying that i) the diffusion times need to be increased relative to brain dMRI to reach long-time limits, ii) neglecting cell radius (i.e., using zero-radius approximations[171]) can lead to inaccurate intra-cellular signal representations.

**Other practical considerations.**

**Order randomization**: For studies including the acquisition of multiple *b*-values, it may be convenient to randomize the order of the acquisition of the DW images, so that blocks of highly DW measurements are not acquired at once. Interspacing weaker and stronger diffusion weighting minimizes the risk of gradients overheating and reduces the duty cycle.

**Interspersed b=0 for drift control**: For high field strengths, the RF energy deposition may cause heating of the sample which can be substantial for short TRs and compromise the temperature stability required for usable dMRI data. A temperature steady-state may be eventually reached but transients will be present in the beginning of the scan. Interspersed *b*=0 images are a very effective way of controlling for these effects, scanner stability and image quality throughout long acquisitions[172]. A slow drift in *b*=0 amplitudes across time can be corrected by using detrending, applied to all diffusion-weighted images. When diffusion data is acquired in multiple experiments, it is also important to ensure that the adjustment parameters (reference power, receiver gain, etc.) are consistent, ideally by preventing them from being updated between scans.

**Full-sphere directional coverage**: It is recommended to optimize the distribution of diffusion-encoding directions, as done for humans dMRI. Directions should be distributed across all shells (i.e. using electrostatic repulsion within and between shells) and should cover the full sphere instead of the half-sphere[139,140]. This coverage is optimal unless there is a specific need to acquire the same directions on each shell, for directional fits of diffusivity and kurtosis for instance. The recommended schemes cannot usually be generated by the vendor software and should be generated separately and imported into the system as a custom gradient direction file

– (see Section 5.1 on open source resources, and https://github.com/ecaruyer/qspace, in particular).

**Effective b-matrix**: It is crucial to use the effective b-matrix for dMRI analysis, rather than the nominal one. The effective b-matrix is typically provided in an output file collecting all acquisition parameters. For non-vendor sequences, it may need to be calculated directly using the sequence diagram and all known gradients played in the sequence. Ice water[173] and other pure liquids, particularly those with low diffusivities (see Table 1 in Fieremans and Lee[174]) can be useful phantoms to assess whether the correct b-matrix is used. Such phantoms can also be used to measure the effect of gradient spatial non-linearity on the effective b-matrix across the whole imaging field, as well as to test for spurious "diffusion time-dependence" from scanner drifts, etc. It is also recommended to include the effective b-matrix when reporting methods, to improve between-site comparisons and thus increase the value of animal study results toward clinical translation and validation.

**Where future work should lie**

Pre-clinical MRI vendors are encouraged to implement diffusion product sequences where the default settings account for the practical considerations mentioned above, and to open the possibility for multi-shell acquisitions with different directions across different shells.

The harmonization of DTI acquisition protocols (e.g. as to the choice of a maximum $b$ = 1000 s/mm$^2$ *in vivo*) may help multi-site reproducibility and comparison studies. Notwithstanding, encouraging the community to acquire richer datasets by default (e.g. multi-shell at minimum, but even for multiple diffusion times) can open up many avenues for testing new models retrospectively on public datasets, in a variety of animal models, healthy and diseased.

The development of biophysical models of tissue should uphold high standards in terms of accuracy and precision of microstructural features estimated, and validated using complementary techniques such as light microscopy.

Finally, the flexibility associated with preclinical MRI scanners will hopefully foster further developments in terms of novel diffusion encoding and acquisition techniques to bring dMRI ever closer to *in vivo* histology.

### 3.6 Spatial resolution

The image spatial resolution is a critical decision in any experimental process. Although brain dimensions vary by orders of magnitude across species from the mouse (0.4mL) to human (1300mL) the relative size of voxels to the size of the brain are generally consistent across many species. Put simply, spatial resolution should be as high as permissible for the targeted SNR and scan time.

Anisotropic resolution, with thicker slices than in-plane voxel size, is the most widespread design because it is fast and less gradient demanding, while yielding higher SNR than isotropic resolution. Depending on the application, isotropic resolution may however be warranted, e.g. for tractography and for interpretation of morphological details and anatomical boundaries. Thick slices will introduce more partial volume effects and can challenge the quality of image registration.

Below we provide typical volumes of brains, and compute what the equivalent voxel size (i.e, the **volume equivalent resolution**) would be given the ratio of volumes, and a typical 2-mm isotropic human scan (**Table 1**). We have chosen 2-mm isotropic as a 'standard' only for comparison purposes, and note that larger voxel sizes (2.5-mm or 3-mm isotropic) are common in dMRI, while smaller voxels are also possible with novel acquisition strategies[175–178] or stronger gradients (<1-mm isotropic). Similar figures hold for other organs used in dMRI literature[179].

| Species | Brain Volume (mL) | Matching spatial resolution (isotropic) | Reported in literature (*in vivo*) | WM/GM volume ratio |
|---|---|---|---|---|
| Human | 1200 | 2 mm | 1 — 3 mm isotropic | ~55/45 |
| Mouse | 0.4 | 140 μm | 100 — 200 μm in-plane<br>200 — 300 μm slice thickness<br>"Extreme" datasets: 75 — 125 μm isotropic | ~8/92 |
| Rat | 1.7 | 225 μm | 100 — 300 μm in-plane<br>250 — 600 μm slice thickness | ~13/87 |
| Squirrel monkey | 35 | 600 μm | 600 — 700 μm in-plane<br>1 mm slice thickness | ~37/63 |
| Macaque | 80 | 800 μm | 500 μm — 1 mm in-plane<br>1 — 2 mm slice thickness<br>"Extreme" datasets: 580 μm isotropic | ~47/53 |

**Table 1**. Summary of brain volumes of various species, matching spatial resolutions to a typical human dMRI, and ranges of *in vivo* spatial resolutions from the literature. A few references are provided, but are not comprehensive. Resolutions reported in literature for human[180], mouse[21,86], rat[165,181], squirrel monkey[182], macaque[22,183,184]. WM/GM volume ratio are reported in Ventura-Antunes et al.[185] where GM volume is a measure of total cortical volume.

## 4. Data Processing

We refer to *pre-processing* as steps that come before any diffusion fitting (tensors, biophysical models, etc.). Pre-processing includes data conversion (e.g. DICOM to NIfTI), noise reduction, artifact correction/mitigation or any step that aims at improving data quality. Processing refers to diffusion data fitting and normalization to standard space.

### 4.1 Pre-processing pipeline

#### 4.1.1 Brain

Generally, pre-processing diffusion datasets of preclinical acquisitions is similar to that of the *in vivo* human brain. Below we detail the steps associated with a typical pre-processing pipeline, stressing in particular what may differentiate *in vivo* small-animal from human implementations, and how available tools can/should be adapted accordingly.

Pre-clinical scanner software often outputs data in their own vendor-specific format, with recent versions offering the possibility to export the data as DICOM or NIfTI directly. Overall, explicit conversion by the user to one of the aforementioned formats - typically using in-house written code - is still widespread, which entails possible incompatibilities with BIDS format, data sharing and processing multi-center data. Some solutions such as DICOMIFIER [186] exist, and the adoption of a standard tool by the community - or by the vendors - will greatly aid data harmonization.

The definition of diffusion gradient directions may not be consistent across vendors or across in-house written data conversion pipelines, with some given in the “imaging frame” i.e. in relation to readout, phase encode, slice direction (or second phase-encode direction) and others in the “lab-frame” with Z being typically along the direction of the main magnetic field. The orientation of the images with respect to the applied diffusion directions is very important, particularly for tractography, and should be checked carefully. Tools are also available to systematically correct any errors[187,188].

Similarly to *in vivo* human diffusion imaging, the recommended data pre-processing pipeline includes: reduction of random noise (“denoising”)[189], Gibbs ringing correction[190–192], combined eddy-current, motion and susceptibility distortion correction[193], along with gradient non-linearity correction (if applicable), Rician bias correction and signal drift correction due to scanner instabilities[172,194]. Recent examples of the concatenation of these processing steps into a pipeline is DESIGNER and PreQual[195,196], or the pre-processing implementation in MRTrix3 [197].

We briefly recapitulate a few common pitfalls and/or quality checks and we refer the reader to the references cited above for comprehensive descriptions of the techniques and their applicability.

The very first prerequisite for most pre-processing steps is providing a **brain mask,** which is a weakness of non-human brain preprocessing pipelines. This can be the consequence of either algorithms using inappropriate priors for non-human brain anatomy in atlas-based brain masking (e.g. very different shapes and/or sizes between human and rodent), or a consequence of bias fields (inhomogeneous $B_1$ receive field for surface coils and potentially transmit field when using surface emitters) which strongly affect the performance of intensity-based brain masking. *Bias field correction* on a *b*=0 image can be performed with a variety of software, but should be used for the sole purpose of brain extraction and not as input for the remainder of the pre-processing and processing pipeline. Dedicated tools to perform brain extraction using *registration to a matching species atlas* are also available (e.g. https://github.com/jlohmeier/atlasBREX[198,199]). If an atlas database for the species and MR contrast of interest is available, the multi-atlas label fusion segmentation approach performs very well[200,201]. More recently, *deep-learning-based frameworks* have been developed and validated to identify the MR image boundaries of the brains of rodents and non-human primates[202–204]. Finally, in the early pre-processing steps the brain mask could be inflated to yield a generous brain inclusion, making sure it encompasses distortions and motion that have not yet been corrected. At the end of the pipeline, a second, more refined brain masking may be performed.

**Denoising** aims at reducing *random noise*, i.e. random fluctuations or Nyquist-Johnson noise (and its digital counterpart), in contrast to physiological noise or any other structured unwanted signal contribution (e.g. cardiac or respiratory fluctuations). A principal component analysis can be used to denoise a 4D stack of diffusion-weighted images[205]. The differentiation between signal- and noise-carrying principal components has recently been automated by adopting principles from Random Matrix Theory[189]. This technique requires that (a) the noise level is constant and uncorrelated across all diffusion-weighted images and (b) that the number of diffusion-weighted images is large - the suggested use is 30 images or more. If such requirements are not met, then alternative denoising strategies are presented[206,207]. Various supervised, unsupervised, and self-supervised machine learning-based techniques have recently been developed and evaluated for the denoising of such data[208]. Inspecting the residuals between the raw and denoised data for absence of structural content is a good quality check.

The implementation of **Gibbs ringing correction**[190,191] has a large positive impact on microstructure model estimates, where the "corruption" of voxels by neighboring CSF can change the apparent microstructure composition dramatically, but has very little impact on the estimation of the fiber orientation distribution for tractography purposes[190,209,210]. One common limitation is the use of partial Fourier for the acquisition, which makes the correction as implemented by Kellner et al.[190] less effective. A new approach for this correction in partial-Fourier data has been recently developed[191]; both these methods are suitable for 2D multi-slice acquisitions.

The 'topup' and 'eddy' tools in FMRIB's Software Library (FSL)[211], which can correct for **susceptibility distortion, eddy current or motion-related jitter** require two specifics of data acquisition that are often not the default on pre-clinical scanners. On the one hand, for susceptibility-distortion correction a few *b*=0 images with reversed phase-encode direction should be acquired to enable the calculation of the distortion field. If these images are not available, alternative distortion correction methods might include nonlinear registration to undistorted anatomical images, or correction using a fieldmap ($B_0$ map). On the other hand, for eddy current correction, 'eddy' requires diffusion directions that are distributed over the entire sphere, and not the half-sphere. If this sampling is not available in the product sequence, a custom diffusion direction file should be provided for the acquisition. Finally, it is important to note that default FSL parameters are suitable for human dMRI, and need to be tuned to the sample of interest in preclinical imaging (e.g. the resolution or knot-spacing of warp fields for topup, or any parameters that are dependent upon spatial scale).

**Rician bias correction** consists of correcting the diffusion signal decay by subtracting the non-zero Rician floor. For software and methods, see[195,212] and Section 5.1.1. One substantial advantage of preclinical MRI data is that coils are often single-channel or quadrature recombined, and the complex-valued data is more easily retrievable from the scanner. Complex-valued data from a single-channel coil is by design characterized by Gaussian noise. Rician bias can be thus minimized by denoising in complex space, and possibly also circumvented entirely by working with real-valued data after phase unwrapping[213].

Finally, **temporal instability** on the scanner due to magnet drift or gradient heating can be measured and corrected by collecting multiple b=0 images throughout the scan (see Section 3.5). Although this is not commonly done in the literature, we advocate for instability correction; methodology and code are described in [172] and in Section 5.1.1.

### 4.1.2 Spinal cord & other organs

Spinal cord dMRI focuses on white matter which is located at the periphery (while the gray matter is inside). Typical preprocessing steps for spinal cord include:

1. **Segmenting** the outer contour of the spinal cord and the gray matter (which also results in white matter mask). This can be achieved using active contour[214], propagation of a 3D mesh[215] or deep learning[216]. The two latter methods are available in the Spinal Cord Toolbox (SCT)[217]. For more details on spinal cord segmentation, please see [218].
2. **Straightening** the spinal cord to have it aligned along the superior-inferior axis. The benefit of this step is to facilitate the registration to a spinal cord template and atlas, and/or for group analysis. Straightening the spinal cord can be done with SCT using an algorithm that preserves the topology of the internal structure of the spinal cord[219].
3. **Registering** the spinal cord to an anatomical template. This step is useful for extracting diffusion metrics within specific white matter tracts of the spinal cord, e.g., cortico-spinal, rubrospinal, dorsal columns. There exist spinal cord templates and atlases, for example a rat spinal cord template[220].

An end-to-end analysis pipeline, with documentation, example data, and procedure for manual correction is available at http://spine-generic.readthedocs.io/ [221]. This project is for *in vivo* human spinal cord, but could be adapted for *ex vivo* and non-human species.

dMRI provides key information on microstructural properties also in other organs such as liver[179], kidneys[222,223], muscle[224] or heart[225]. Imaging each of these anatomical districts *in vivo* comes with its own challenges, mainly related to complex motion patterns due to proximity to the lungs, inhomogeneous magnetic fields close to air cavities (e.g., stomach, lungs, rectum), pulsation effects, intrinsic low signal-to-noise ratio due to short $T_2$ (e.g., liver iron). Typical pre-processing steps used for brain MRI can also be useful in these anatomical areas, e.g., denoising, motion and distortion correction, although at present there is still a lack of processing packages tailored for these applications.

**Where future work should lie**

Brain extraction is an important preprocessing step which, if inaccurate, can largely affect the performance of downstream steps. Preprocessing tools, at each step of the pipeline, should account for geometric and anatomical differences between species, and studies should be performed to optimize and standardize these tools depending on species. Finally, setting up a publicly available pipeline that integrates these pre-processing steps seamlessly, with optimized parameters for each species would be highly beneficial.

## 4.2 Processing pipelines

After pre-processing, it is typical to perform voxel-wise analysis of the diffusion-weighted measures to output parametric maps of a variety of derived metrics. These parametric maps can undergo subsequent analysis at the region of interest (ROI), individual, or group levels.

DTI, DKI, or biophysical models suited for the tissue of interest (white matter, gray matter, muscle, various tumor types, etc.) can be applied to small-animal data [6,7,161,226,227]. For DTI, many software tools are available - see Section 5.1. Substantially fewer software packages offer diffusion kurtosis estimation [197] and most software do not check for *b*-value suitability prior to DTI or DKI estimation. For biophysical model estimation, dedicated code is usually provided by the model developers.

Once parametric maps of various diffusion metrics are available in native space, it is common to use registration either to import atlas-based segmentation of brain regions for ROI analysis or to bring individual maps into a common space for voxel-based comparisons. For this registration/normalization step, typical tools used in human data also work well for animal data, with some customization. For non-linear registration for instance, default physical dimensions of warp and smoothing kernels should be scaled to those of small-animal brains.

**Where future work should lie**

Similar to preprocessing, free online sharing of processing tools can accelerate the harmonization process, with several efforts going in this direction (see also Section 5.1.1). Moreover, prospective harmonization studies[228] are required to understand and account for inter-site variability.

## 4.3 Tractography

The application and use of dMRI-based fiber tractography to study fiber pathways and wiring diagrams of the brain remain largely the same for small animals as for the *in vivo* human[229]. The fundamentals of tractography (deterministic and/or probabilistic algorithms) also remain the same. In our experience, the only required change is the brain masking approach (see Section 4.1.1).

**Fiber Orientation Estimation**: Minimal changes are needed in voxel-wise reconstruction steps and approaches that estimate fiber orientation distributions will work adequately in animal model systems (including diffusion tensor imaging, spherical deconvolution, ball & sticks, and more advanced methods), resulting in a field of orientation estimates that can be used for tractography.

**Tractography**: Several changes are needed in the tractography process itself. For example, with smaller brains and smaller voxel sizes, it is common that the “step size” in tractography algorithms must be reduced, although this will usually be performed by default in many software packages (e.g. MRtrix3, DSI Studio, DIPY, FSL, ExploreDTI). Often, false positive streamlines are removed through filtering or clustering operations. If thresholding by streamline length (i.e., setting a minimum or maximum length to eliminate spurious or implausibly long streamlines), these thresholds must be adapted to an appropriate measure for each specimen and the pathway or system under investigation. The process of filtering by microstructural measures, or the diffusion signal (using algorithms such as LIFE[158], SIFT[156,157], or COMMIT[155]), as well as adding anatomical constraints, is also relevant in animal models.

One tractography application is **bundle segmentation**. Individual fiber pathways, or fascicles, of the brain are virtually dissected to be studied across cohorts or time. This is usually done by using ROIs through which bundles must or must not pass, to isolate a desired pathway. The most obvious change for animal models is to consider the species-specific brain anatomy, and ROIs must be modified in accordance with prior knowledge of the bundle[230–232]. For this reason, many automated or manual protocols for bundle dissection are not readily adaptable for small animals. Although some works have described and created tools for dissection in some species (for example macaque[233], squirrel monkey[234], mouse[235]), common tools and protocols for bundle segmentation in humans[236,237] have not yet been adapted. Regarding the analysis of bundles, quantifying microstructure along or within the bundle of interest[238 239], connectivity and shape of the bundles[240] can be done using the same analysis as for human data.

The next common application of tractography is **connectome analysis** - an analysis of the set of streamlines throughout the entire brain to determine network properties, often using graph-theoretic measures. Potential differences in connectome analysis include different edges/nodes used to derive the matrix, which will typically be derived from existing templates and atlases[234,241–244].

**Where future work should lie**

While the process of tractography in small animals strongly parallels that of human, there are still areas for which no guidelines can be provided. For example, it is unknown what spatial resolution is necessary/optimal in various animal brains.

For the tractography process, it is unclear if specific modifications to the generation of streamlines are needed. For example, it may be necessary to adapt starting/stopping criteria, curvature thresholds, or anisotropy criteria for animals with different GM/WM volumes, different (both more or less complex) geometries, and different expected curvatures or pathways. Additionally, with the benefits of strong fields and high gradients on preclinical magnets comes the ability to image, and possibly probe connections within cortical or deep gray matter areas, an area that is relatively unexplored in the human brain, and for which there are little-to-no guidelines.

Also, few guidelines are provided on optimal regions and region placement for bundle segmentation. Many human analyses rely on regions located in a standard space (typically MNI 152), that have been modified and tailored over the years to continually improve the resulting tractography dissections. Few studies exist which describe appropriate region placement in animal models due to a sparsity of resources dedicated to tractography in the animal models. Future work should lie in creating resources that allow whole brain tractography (possibly informed by anatomical constraints) in various models, followed by atlas-based labeling (to create nodes/edges for connectome generation), and bundle dissection for pathways of interest.

Tractography is often validated in animal models but we should strive to understand and quantify differences between tractography based measures of tissue orientation, and experimental tract-tracing methods that visualize specific neural connections.For example, knowledge of complications due to crossing fibers[245], due to spatial resolution[246,247], bottle-neck regions of tractography[64,248,249], superficial U-fibers[250], effects of experimental parameters[251], and false-positives and false-negatives[71] have been elucidated through the use of animal models - and we should continually understand the parallels to the human brain to reliably interpret measures derived from tractography. We also urge investigators to fully document the tractography parameters in their publications for rigor and reproducibility.

### 4.4 Group-level analysis

A variety of group-level analyses are available for preclinical dMRI. The choice between them will be guided by the statistical power (number of animals available vs expected effect size) and by the study design, whether it is hypothesis-driven or exploratory.

ROI analyses consist in computing the mean (or median) of a given dMRI metric in an anatomical ROI and testing statistical differences of the ROI-level values in the variable of interest (groups, timepoints, conditions, etc.). Ideally, to limit bias, the ROI segmentation is performed using registration to an atlas where the structure of interest is segmented (and the segmentation is then propagated back to individual space using the inverse transforms). A good alternative is to register all datasets to a study-specific template. The ROI can be manually drawn on the template and then also propagated back to individual space. For highly deformable or heterogeneous structures (e.g. viscera or tumors) manual ROI segmentation on each dataset may be unavoidable. Operator bias should be carefully mitigated in such cases. ROI segmentations should always be visually inspected for accuracy and consistency. Corrections for multiple comparisons should be applied if multiple ROIs are considered. ROI-based group analyses are suitable for low-powered studies (few animals, weak effect) or hypothesis-driven studies (the ROI where the effect is expected is known a priori). However, if the effect is very localized, large anatomical ROIs may reduce the power by averaging across voxels where no effect is to be expected.

Voxel-based analyses consist in registering all quantitative dMRI maps to a common template and identifying clusters of voxels that display a significant difference in the variable of interest. For brain studies, a common tool for this effect is FSL's TBSS[239]. Voxel-based analyses are suitable for exploratory studies and for highly-powered studies (large cohorts or large effect) due to the strict statistical threshold when correcting for many comparisons (large number of voxels), typically using permutations tests.

Analysis, interpretation, comparison across studies, and re-use of data depends on specific attribution and The documentation of anatomical location is important for data interpretation, re-use and comparison across studies, and can be communicated using spatial coordinates or anatomical terms [252,253]. Several open access 3-D reference brain atlases are available for different species (see Section 5.1.2).

## 5 Perspectives

### 5.1 Open science

#### 5.1.1 Code/Software

Challenges with pre-processing and processing pipelines highlighted above can be overcome through code sharing and harmonization of implementations.

In order to allow for a more dynamic and self-updating resource center, and facilitate code sharing, we have compiled a (non-comprehensive) list of available software dedicated to the acquisition and processing of preclinical dMRI data, meant to be updated regularly, on a public repository: https://github.com/Diffusion-MRI/awesome-preclinical-diffusion-mri.

Updates on available software and tools can be shared between developers and users.

### 5.1.2 Data Sharing & Databases

A critical aspect of data sharing is that data should be findable, accessible, interoperable, and reusable, in 2016 formulated as the F.A.I.R principles[243]. These principles are now widely adopted by researchers, universities, funding agencies, and journals.

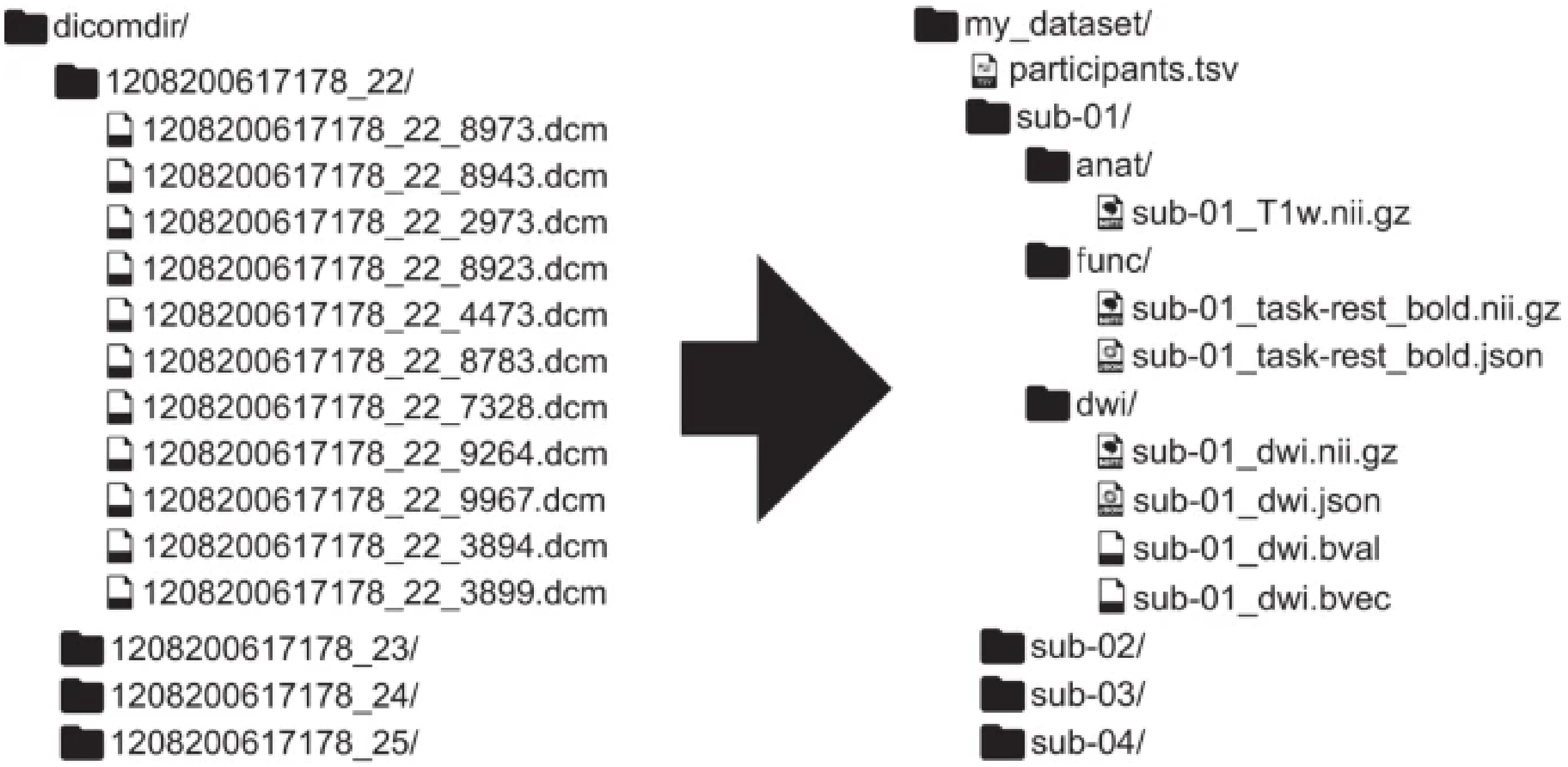

**Figure 5**. Illustration of a BIDS structured dataset (right) starting from vendor-specific convention of data organization (left). From [255]

Standards for naming and organizing folders and files are of key importance for the reusability of shared imaging data. The neuroimaging community therefore proposed Brain Imaging Data Structure (BIDS)[255], recently endorsed as a standard by the International Neuroinformatics Coordinating Facility (INCF)[256]. In brief, using BIDS, data are organized according to contrasts (*anat*, *dwi*). File names include relevant suffixes that help researchers and software to identify origin and intention of the files (e.g., "_dwi" is intended for diffusion weighted analysis). Sidecar JSON files include additional metadata that are relevant for the analysis. **Figure 5** illustrates a dataset structured according to BIDS. While the BIDS standard has originally been motivated by the brain functional MRI community, this standard is being actively expanded to accommodate more MR techniques and modalities, the latest of which also include advanced diffusion-weighted imaging[257]. Also relevant are recent initiatives for incorporating animal data into the BIDS standard (https://bids.neuroimaging.io/bep032). In the perspective of increasing data sharing opportunities, as well as generating traceable and comparable datasets, we strongly encourage organizing raw pre-clinical data using the BIDS format. In addition to facilitating data sharing, data organization standards like BIDS help designing applications that know *where* to look for input data. This ultimately helps automating analysis tasks and creating pipelines. A list of BIDS-compatible apps is available at https://bids-apps.neuroimaging.io/apps/.

Platforms that could serve as a repository for pre-clinical datasets include OSF (https://osf.io/), OpenNeuro (https://openneuro.org/), Zenodo (https://zenodo.org/), NITRC (https://www.nitrc.org/) and others (https://odc-tbi.org/). The EBRAINS research infrastructure for brain-related research (http://ebrains.eu/) offers solutions for sharing curated data sets with standardized metadata, and links to brain atlases, analytic tools, and solutions for computational modeling and simulation.

To exemplify sharing and reuse of useful imaging data, we compiled a (non-comprehensive) list of links to selected publicly available small-animal or *ex vivo* diffusion-weighted datasets: https://github.com/Diffusion-MRI/awesome-preclinical-diffusion-mri.git.

**Where future work should lie**

Code can be hosted on platforms such as GitHub, GitLab, Zenodo, NITRC etc. Hosting code via these tools is not only beneficial for the community, but also for the code developers themselves (and their respective research groups). Indeed, this ensures code safekeeping, retrieving and versioning. Nevertheless, code sharing and submission to hosting platforms comes with the responsibilities of documenting, cleaning, packaging, testing and versioning the code. These duties come at a (high) cost of requiring an in situ software engineer. Initiatives aimed at allocating special resources for software maintenance via funding bodies are urgently needed.

For licensing open source code, most permissive licenses include MIT and BSD licenses. It means that the code can be reused by any entity (person or company), and importantly to note, is that the modified code can be distributed as closed source. If you wish to enforce the disclosure of your open source code, there are so-called 'copyleft' licenses, such as the GNU GPLv3 and the Mozilla Public License 2.0. For more details, see https://choosealicense.com/licenses/.

Certainly one of the highest aims is to propose a successful, transparent and comprehensive analysis framework that promotes reproducibility.

The amount of open-source code and data is overwhelming. Sharing code and data is a double-edged sword. Indeed, public sharing of scientific objects that do not meet certain standards or requirements can do more harm than good. It is crucial to keep code on a dedicated platform (e.g. GitHub) and point to a specific tag or commit hashtag directly in the associated paper and OSF data repository. In parallel, it is important to version-track the dataset itself and mention specific versions where appropriate. Several software solutions exist to link data objects and code and track provenance, for example Datalad (https://www.datalad.org/) and the YODA framework (https://handbook.datalad.org/en/latest/basics/101-127-yoda.html).

## 5.2 The future: what should we strive to achieve?

As a field, we should continually strive to achieve reduced barriers to entry for new imaging centers, new scientists, and new industries who aim to use dMRI in a preclinical setting. Towards this end, as a community, we should promote dissemination of knowledge, code, and datasets to achieve high standards of data quality and analysis, reproducibility and transparency. We should foster academic and industrial collaborations with MR vendors, as well as reduce globally the time and cost of research in this field.

## Acknowledgements

The authors acknowledge financial support from: the National Institutes of Health (K01EB032898, R01AG057991, R01NS125020, R01EB017230, R01EB019980, R01EB031954, R01CA160620, R01NS109090, R01NS119605, U54AG054349, P50MH096889), the National Institute of Biomedical Imaging and Bioengineering (R01EB031765, R56EB031765), the National Institute on Drug Abuse (P30DA048742), the Secretary of Universities and Research (Government of Catalonia) Beatriu de Pinós postdoctoral fellowship (2020 BP 00117), “la Caixa” Foundation (ID 100010434), the European Union's Horizon 2020 research and innovation programme under the Marie Skłodowska-Curie grant agreement No 847648, Junior Leader fellowship codes (LCF/BQ/PR22/11920010; CF/BQ/PI20/11760029), The European Union’s Horizon Europe Programme for Research Infrastructures under the Specific Grant Agreement No.101147319 (EBRAINS 2.0), the Research Council of Norway under grant agreement no. 333157 (INCF Norwegian Node 2022-2027), the Research Foundation Flanders (FWO: 12M3119N), the Belgian Science Policy Prodex (Grant ISLRA 2009–1062), the μNEURO Research Center of Excellence of the University of Antwerp, the Institutional research chair in Neuroinformatics (Sherbrooke, Canada), the NSERC Discovery Grant, the European Research Council Consolidator grant (101044180), the Canada Research Chair in Quantitative Magnetic Resonance Imaging [950-230815], the Canadian Institute of Health Research [CIHR FDN-143263], the Canada Foundation for Innovation [32454, 34824], the Fonds de Recherche du Québec - Santé [322736], the Natural Sciences and Engineering Research Council of Canada [RGPIN-2019-07244], the Canada First Research Excellence Fund (IVADO and TransMedTech), the Courtois NeuroMod project, the Quebec BioImaging Network [5886, 35450], the Mila - Tech Transfer Funding Program and the Swiss National Science Foundation (Eccellenza Fellowship PCEFP2_194260).